\documentclass[aps,prl,superscriptaddress]{revtex4-1}

\usepackage{graphicx}
\usepackage{amsmath}
\usepackage{color}

\begin{document}

\newcommand*\rfrac[2]{{}^{#1}\!/_{#2}}

\title{Self-tracking Energy Transfer for Neural Stimulation in Untethered Mice} 



\author{John S. Ho}
\affiliation{Department of Electrical Engineering, Stanford University, Stanford, California 94305, USA}
\author{Yuji Tanabe}
\affiliation{Department of Electrical Engineering, Stanford University, Stanford, California 94305, USA}
\author{Shrivats Mohan Iyer}
\affiliation{Department of Bioengineering, Stanford University, Stanford, California 94305, USA}
\author{Amelia J. Christensen}
\affiliation{Department of Bioengineering, Stanford University, Stanford, California 94305, USA}
\author{Logan Grosenick}
\affiliation{Department of Bioengineering, Stanford University, Stanford, California 94305, USA}
\affiliation{Neurosciences Program, Stanford University, Stanford, California 94305, USA}
\author{Karl Deisseroth}
\affiliation{Department of Bioengineering, Stanford University, Stanford, California 94305, USA}
\affiliation{Neurosciences Program, Stanford University, Stanford, California 94305, USA}
\affiliation{Department of Psychiatry and Behavioral Sciences, Stanford University, Stanford, California 94305, USA}
\affiliation{Howard Hughes Medical Institute, Stanford University, Stanford, California 94305, USA}
\affiliation{CNC Program, Stanford University, Stanford, California 94305, USA}
\author{Scott L. Delp}
\affiliation{Department of Bioengineering, Stanford University, Stanford, California 94305, USA}
\affiliation{Department of Mechanical Engineering, Stanford University, Stanford, California 94305, USA}
\author{Ada S. Y. Poon}
\affiliation{Department of Electrical Engineering, Stanford University, Stanford, California 94305, USA}


\date{\today}

\begin{abstract}
Optical or electrical stimulation of neural circuits in mice during natural behavior is an important paradigm for studying brain function. Conventional systems for optogenetics and electrical microstimulation require tethers or large head-mounted devices that disrupt animal behavior. We report a method for wireless powering of small-scale implanted devices based on the strong localization of energy that occurs during resonant interaction between a radio-frequency cavity and intrinsic modes in mice. The system features self-tracking over a wide (16~cm diameter) operational area, and is used to demonstrate wireless activation of cortical neurons with miniaturized stimulators (10~mm$^{3}$, 20~mg) fully implanted under the skin.
\end{abstract}


\maketitle 

Direct manipulation of neural circuits in small animal subjects (such as mice) is key to understanding brain function because controlled perturbation of neural activity during natural behavior is needed to establish the causal role of specific neural states in behavioral outcomes~\cite{Deisseroth_2014}. In many neuroscience experiments, neural activity is manipulated either electrically or optically using wires or optical fibers inserted into the brain~\cite{Aravanis_2007,Holtzheimer_2011,Warden_2014}. Such tethered systems, however, do not permit free movement, protrude beyond the skin, and limit the environments in which experiments can be performed.

Wireless systems for neural modulation have been developed in order to eliminate tethers~\cite{Arfin_2009,Szuts_2011,Wentz_2011,Kim_2013,McCall_2013,Yeh_2013,Jow_2014,Hashimoto_2014}. Most systems for mice need to be wirelessly powered due to the substantial size and weight of batteries relative to that of the animal~\cite{Talwar_2002,Szuts_2011}. Using resonant coils, near-field energy transfer has been demonstrated for powering devices worn on the head for both wireless stimulation and recording~\cite{Wentz_2011,Jow_2014}. Head-mounted antenna modules have also been developed for harvesting far-field radiation in order to power microscale LEDs inserted through a flexible membrane~\cite{Kim_2013,McCall_2013}. Although these systems enable free motion, they remain limited by the size of the power harvesting structures on the device, which all have been so far head-mounted. While electronic or photonic components can be miniaturized, the dimensions of the powering structure is constrained by the energy transfer performance. The transfer efficiency should, in typical applications, exceed $10^{-4}$ (considering device power levels of 500 $\mu$W at a maximum output power of $5$~W for safety), while the coverage area should be greater than 10~cm in diameter in order to permit free movement. These stringent requirements have so far prevented the use of devices with dimensions on the order of 1~mm that can be fully implanted under the skin.

Here we report a method for wireless energy transfer that achieves a wide area of coverage and high transfer efficiency (on the order of $10^{-3}$) to miniaturized coils (2~mm in diameter) that can be fully implanted in mice. We show that by coupling a radio-frequency (rf) resonator to the tissue volume, millimeter-scale devices can be powered over a large circular region (16~cm diameter), without any tracking mechanisms. Power sufficient for most neuromodulation tasks can be delivered under low levels of rf exposure~($<$4~W averaged). As a proof-of-concept, we use the system to power an electrostimulation device that activates the neocortex of a behaving mouse.

As dielectric objects, mice support a series of electromagnetic modes that give rise to strong localization of energy. By tailoring the properties of a metallic resonator to couple to these modes, efficient energy transfer can be realized. Our powering system is based on a cylindrical radio-frequency (rf) resonant cavity  [Fig.~1(a)] patterned with subwavelength apertures~\cite{Yang_2002,Wen_2005}. Energy can be transferred from the cavity resonance to an external dielectric object, such as a mouse, through the fringing evanescent field [Fig.~1(b)]. The operation of the system can be described using coupled-mode theory~\cite{Kurs_2007,Yu_2011,Kim_2013b}. Consider the fundamental modes of the cavity and object with electric field patterns $\mathbf{e}_{1}$ and $\mathbf{e}_{2}$ respectively. The total electric field can be approximated as $a_1(t)\mathbf{e}_{1}(\mathbf{r}) + a_2(t)\mathbf{e}_{2}(\mathbf{r})$ where $\mathbf{e}_{1}$ and $\mathbf{e}_{2}$ are normalized such that $|a_1|^2$ and $|a_2|^2$ are the energy in the modes. The evolution of the mode amplitudes are described by the equations
\begin{equation}
\frac{d}{dt}\left(
\begin{array}{c}
a_{1} \\
a_{2}
\end{array}
\right) = \left(
\begin{array}{cc}
-i\omega_{1}-\gamma_{1} & \kappa \\
\kappa & -i\omega_{2}-\gamma_{2} \\
\end{array}
\right) \left(
\begin{array}{c}
a_{1} \\
a_{2}
\end{array}
\right)
\label{eq:coupled_mode}
\end{equation}
where $\kappa$ is the coupling coefficient, $\omega_{n}$ the resonant frequencies, and $\gamma_n$ are the intrinsic decay rates. The cavity is driven by a continuous wave source at frequency $\omega$ such that the steady-state amplitudes have a time dependency $e^{-i\omega t}$.  This frequency is set to $\omega=\omega_{1}$ in order to maximize coupling between the continuous wave source and the cavity resonator. From energy considerations, the parameter $\kappa$ can be expressed as
\begin{equation}
\kappa(\mathbf{r}_{0}) = \frac{i\omega}{4}\int d^{3}r~ [\epsilon(\mathbf{r}) - \epsilon_0]  \mathbf{e}_{1}^{*}(\mathbf{r})  \cdot \mathbf{e}_2(\mathbf{r}-\mathbf{r}_{0})
\end{equation}
where $\mathbf{r}_{0}$ is the position of the object on the surface of the cavity and $\epsilon$ the dielectric permittivity of the object. From Eq.~(2), it can be shown that the fraction of power transferred from the cavity to the dielectric object is
\begin{equation}
\eta_{12} = \frac{\frac{|\kappa|^2}{\gamma_{1}\gamma_{2}}}{1+ \left(\frac{\omega_{1}-\omega_{2}}{\gamma_{2}}\right)^2+\frac{|\kappa|^2}{\gamma_{1}\gamma_{2}}}
\end{equation}
The efficiency is maximized when the resonant frequency of the cavity is set to $\omega_{1}=\omega_{2}$. For cylindrical geometries, this resonance occurs when the ratio of the cross-sectional width and the characteristic size of the object is on the order of the dielectric contrast $\sqrt{\epsilon/\epsilon_{0}}$. Owing to the high dielectric contrast between free-space and biological tissue, excitation of the resonator is expected to lead to strong localization of energy. We analyze the modes supported in mice in order to establish the location of these resonances.

The resonant modes of an object with dimensions and material properties similar to a mouse can be identified under plane wave illumination. The object of study consists of an ellipsoidal ``mouse'' with uniform dielectric permittivity set to muscle tissue (e.g.~$\epsilon/\epsilon_{0}=54+i14.2$ at 1.5~GHz) with Debye dispersion~\cite{Gabriel_1996}. As the frequency and polarization of the incident wave varies, coupling to the distinct modes supported by the object gives rise to peaks in the stored energy spectrum. This stored energy, otherwise dissipated as heat, can be utilized to power electronic devices. Fig.~2 shows that an object of characteristic length 5~cm supports two fundamental modes closely spaced at 1.3 and 1.45~GHz (or wavelengths 3.1~cm and 2.8~cm in tissue), corresponding to incident electric polarizations aligned along the major and minor axes respectively. The peaks redshift as the object dimensions volumetrically increase~[Fig~2(a), (b)]. Higher order modes are observed at frequencies beyond the fundamental modes, characterized by a magnetic field intensity distribution with a centered antinode and maxima along the axis orthogonal to the incident electric polarization. In contrast, coupling to the fundamental modes localizes energy in the center of the object~[Fig.~2(c), (d)]. Losses in the object govern the width of the resonance line shape. The quality factor of resonance in the fundamental modes are estimated to be about 4, with losses dominated by tissue absorption. The broadband nature of the resonance allows for substantial coupling to occur over a broad range of frequencies and object geometries.

Based on these identified modes, we set the dimensions of the cavity to support a resonance at 1.5~GHz when coupled to a continuous-wave source. The resonant frequency can be shifted by adjusting the length of tuning rods inserted in the cavity. Due to the significant overlap between the spectral lines of the object's two fundamental modes, the cavity resonator can simultaneously couple to both modes if the fields are circularly polarized. We implement circular polarization by exciting two degenerate cylindrical modes in the cavity along orthogonal axes with a $\pi/2$ phase difference. This results in an overall coupling efficiency that is independent of the orientation of the object in the surface plane~\cite{Yeh_2013}. For an object of length 5~cm along the major axis, simulations show that the energy is transferred across the lattice with efficiencies from 5 to 10\%.

A key property of the system is self-tracking: energy is localized in the object at all positions on the lattice. This property requires that the position-dependence of $\eta_{12}$ be eliminated by designing a surface with a uniform $|\kappa(\mathbf{r})|$ profile. To achieve uniformity, we set the dimensions of the hexagonal apertures to be much smaller than the size of the object, such that the sum of the overlap between the modal field distribution and the fringing evanescent field in Eq.~(2) remains relatively constant over the surface. Fig.~3(a) illustrates self-tracking to a cylindrical object (4~cm wide and 3~cm high) over this lattice design. The $|\kappa(\mathbf{r})|$ profile decreases by less than 20\% at the edge of the center region 16~cm in diameter~[Fig.~3(b)]. The coupling drops rapidly beyond this region, as expected from the low field overlap near the edge. We experimentally demonstrated self-tracking using a cylindrical container of the same dimensions. The container is filled with a 0.5\% saline solution mimicking the dielectric response of tissue. While varying the position of the object, the magnetic field intensity was recorded inside and outside the container in the plane 1.5~cm above the surface. The magnetic field along the axis, normalized to the incident field recorded in absence of the object, is shown in Fig.~3(c). Both at the center and near the edge of the region, energy is strongly localized in the object with differences in intensity that is in good agreement with the simulated $|\kappa(\mathbf{r})|$ profile. This property enables energy transfer to ``follow'' the object over any path of motion on the surface.

The strong localization of energy within a mouse at resonance enables substantial power to be harvested by millimeter-scale structures. Energy extraction by a coil can be described by the coupled-mode equation 
\begin{equation}
\frac{da_{C}}{dt}=-(i\omega_{C}-\gamma_{C}-\gamma_{L}) a_{C} + \kappa_{C}a_{2}
\end{equation}
where $\omega_{C}$ is the resonant frequency of the coil and $a_{C}$ the amplitude of the magnetic dipole moment $\mathbf{m}_{C}$~\cite{Urzhumov_2011}, normalized such that $|a_{C}|^2$ is the energy in the coil. $\gamma_{C}$ is the intrinsic loss of the implanted coil and $\gamma_{L}$ the rate of energy extracted by the load. The coupling parameter $\kappa_{C}$ is given by
\begin{equation}
\kappa_{C} = \frac{i\omega}{4} \mathbf{b}_{2}(\mathbf{r}_{C})\cdot\mathbf{m}_{C}
\end{equation}
where $\mathbf{r}_{C}$ is the position of the dipole moment and $\mathbf{b}_{2}$ is the normalized magnetic field pattern of the fundamental mode. The efficiency at which power is transferred from the object to the coil is
\begin{equation}
\eta_{2L}=\frac{\frac{|\kappa_{C}|^2}{\gamma_2\gamma_L}}{ \left(\frac{\omega_{C}-\omega_2}{\gamma_L}\right)^2 + \left(1 + \frac{\gamma_{C}}{\gamma_{L}} \right)^2}
\end{equation}
where the decay rate $\gamma_{2}$ has been modified to include energy dissipated in the device. Eq.~(6) implies that power transfer to the load is maximized when $\omega_{C}=\omega_{2}$ and $\gamma_{L}=\gamma_{C}$~\cite{Yu_2011}. These impedance-matching conditions can be met by resonating the coil with a capacitor and choosing a load value equal to the real part of the self-impedance of the coil. Simulations show that a 2-mm diameter coil implanted in the body under the skin extracts on the order of 1\% of the energy stored in the mode under perfect matching conditions.

We experimentally demonstrate energy transfer to a miniature electronic device implanted subcutaneously in the back of a mouse. The device is powered by a three-turn coil that, together with a surface-mount capacitor (4~pF), supports a resonance at 1.5~GHz in a tissue environment. The load consists of a four diode half-wave rectifier, an integrated charge pump, and a light-emitting diode (LED)~[Fig.~4(a)]. Based on the circuitry, the LED pulses at a rate that is a function of the rectified power, enabling the transferred power to be optically measured through the skin of the mouse~[Fig.~4(c)]. Pulses were recorded by video over two 30-s intervals while the mouse was allowed to freely move on a thin plastic floor placed over the lattice. The path of motion over each interval was determined from the position of light pulse. Fig.~4(e) shows the efficiency as a function of time calculated from the spacing between adjacent pulses. The average transfer efficiency over both paths was measured to be~$\eta=0.75\times 10^{-3}$, excluding losses in the circuitry. For a 20\% duty cycle stimulation protocol, this translates to 15~mW of available power at a time-averaged input power level of 4~W. This performance level well exceeds requirements for diverse neurostimulation devices~\cite{Montgomery_2015}.


To demonstrate the utility of the powering system for implantable neural stimulators, we wirelessly powered a device for electrical microstimulation of mouse infralimbic cortex, a potentially clinically relevant cortical region implicated in animal models of depression and anxiety~[Fig.~5(a)]~\cite{Price_2009,Holtzheimer_2011}. The device uses the same coil structure in Fig~4(a), but with a load consisting of a half-wave rectifier directly connected to thin magnet wire (34-gauge) electrodes [Fig.~5(b)], and can be entirely constructed without specialized equipment. The volume (10~mm$^{3}$) and mass (20~mg) is about two orders of magnitude less than previously reported systems, which weigh 0.7 to 3~g~\cite{Arfin_2009,Wentz_2011,Kim_2013}. Implantation was performed using a standard stereotaxic surgical apparatus. After lowering the electrodes into the brain, the device was cemented into the skull and the skin sutured over the device such that it is completely subcutaneous. We placed mice over the cavity and applied 10~min of stimulation (20~Hz, 20~ms) at a peak input power of 4~W to the cavity. Following stimulation, the brain was histologically examined for c-Fos expression (Rabbit anti c-Fos, Abcam ab53036, 1:500), which indicates neuron firing~\cite{Sagar_1988}. Fig.~5(c) shows greatly increased c-Fos activation in neurons around the electrode track [Fig.~5(c), (d)]. Control mice, implanted with non-functional devices, incurred similar electrode-related scarring, but minimal activation of c-Fos in or around the track [Fig.~5(e), (f)]. These results indicate robust wireless activation of neural circuits with fully implanted devices over the operation region.

When the mouse is positioned at the center of the lattice, simulations show that the peak exposure is 4.8~W/kg at an output power of 4~W. This level falls below thresholds for human safety (10~W/kg, averaged over 10~g of tissue)~\cite{IEEE_SAR_2005}. Thermal imaging shows that the heating attributable to rf exposure at these levels is 0.5~$^{\circ}$C over 8~min of continuous input (see Ref.~\cite{Montgomery_2015}). The input power level and the resulting exposure levels can be substantially reduced if low-power device designs or short duty cycle protocols are used. Due to the non-radiative mode of transfer, exposure to the experimenter during normal operation is minimal compared to systems based on far-field transfer.

We have demonstrated a wireless powering system based on the resonant interaction between a rf cavity and intrinsic modes in mice. The strong localization of energy that results from the interaction allows millimeter-scale devices, small enough to be fully implanted in mice, to be powered. The transfer is self-tracking over a large area of coverage, enabling unconstrained motion and natural behavior. Combined with electrode-based or optogenetic technologies for neuromodulation, this methodology may pave the way for new tools for studying the neural basis of disease and health.

\textbf{Acknowledgements.} We thank A. J. Yeh and V. Tsao for assisting in the power measurement experiment.

\newpage

\begin{figure*}[tb]
\centering
\includegraphics[width=8cm]{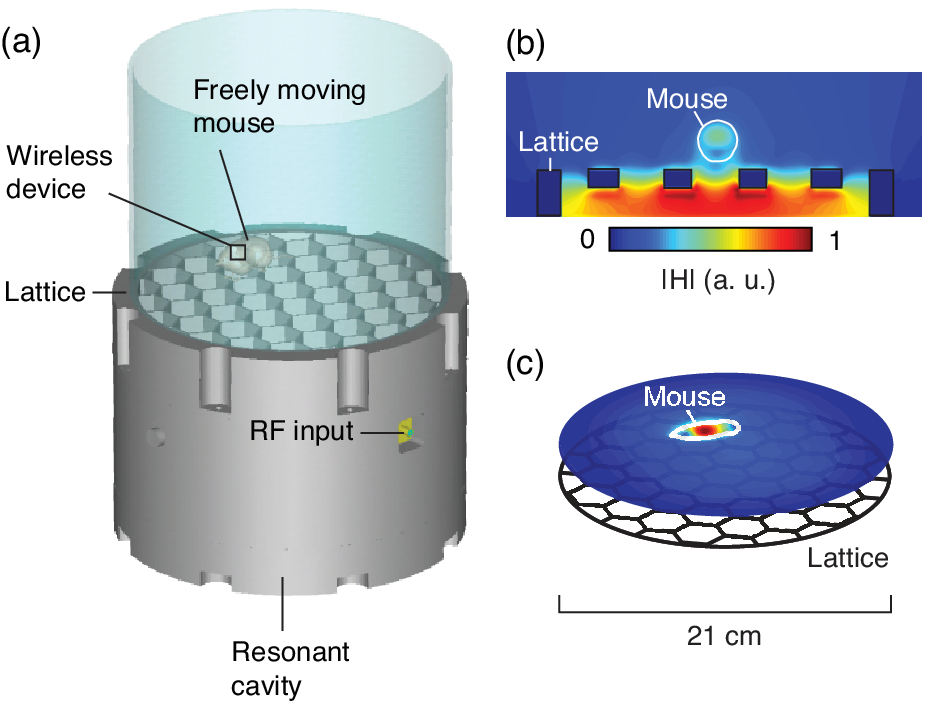}
\caption{Self-tracking energy transfer. (a) Schematic illustrating the operation of the system. The mouse moves freely above the lattice while the resonant cavity is excited with a continuous-wave input. (b) Magnetic field distribution illustrating coupling between the cavity resonance and an intrinsic mode in the mouse. (b) Magnetic field distribution 1.5~cm above the surface at 1.5~GHz. The diameter of the cavity is 21~cm.}
\end{figure*}

\begin{figure*}[tb]
\centering
\includegraphics[width=14cm]{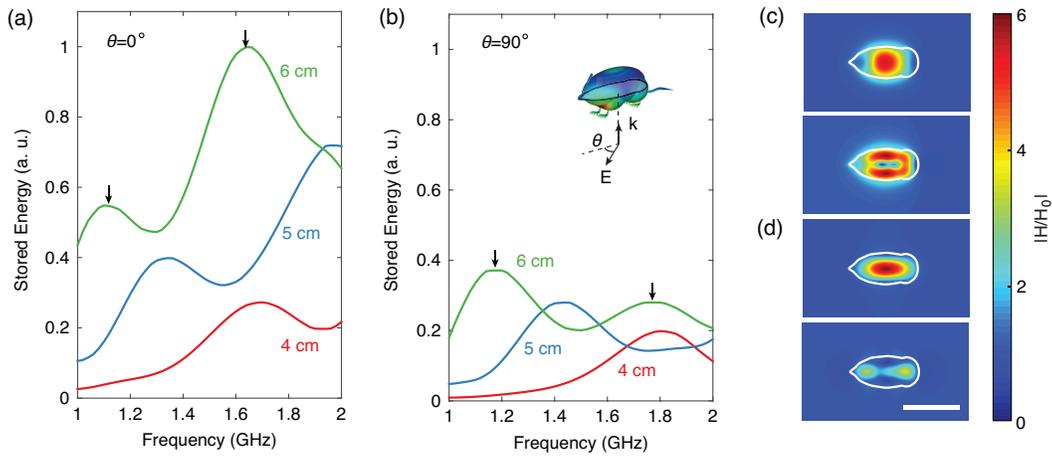}
\caption{Characterization of modes under plane wave illumination. (a), (b) Energy stored in the object as a function of frequency when the incident electric field is parallel to (a) the major axis and (b) the minor axis of the object for different object sizes: major axis length 4, 5, and 6~cm (uniform scaling). (c), (d) Contour plot of the magnetic field distribution corresponding to the peaks indicated by the arrows in (a) and (b) respectively. The field intensity is normalized to the incident field intensity $|H_{0}|$. Scale bar: 5~cm.}
\end{figure*}

\begin{figure*}[b]
\centering
\includegraphics[width=8cm]{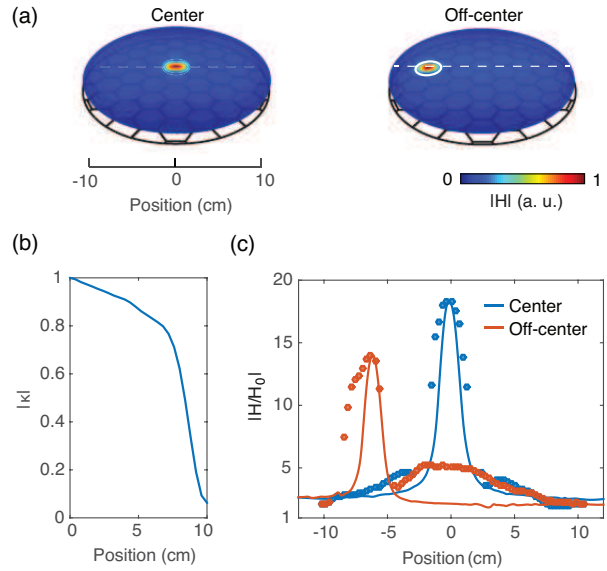}
\caption{Demonstration of self-tracking energy transfer to dielectric object. (a) Contour plot of the field profile above the lattice for a centered and off-centered dielectric sphere (diameter 2.8~cm). (b) Calculated $|\kappa|$ as a function of distance from the center. (c) Measured magnetic field profile (dots) and simulation fit (solid lines) along the center axis. The magnetic field is normalized to $|H_{0}|$, the field recorded in absence of the object.}
\end{figure*}

\begin{figure*}[b]
\centering
\includegraphics[width=14cm]{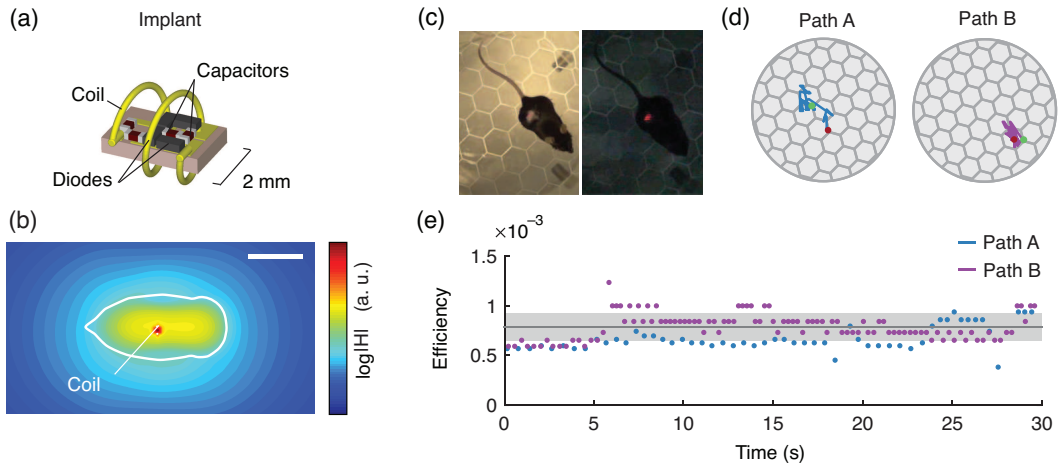}
\caption{Energy transfer to a power measurement device. (a) Schematic of the device consisting of a 2-mm diameter coil, capacitors, diodes, an integrated circuit, and a LED. (b) Simulated magnetic field distribution for a device implanted in the back of a mouse. Scale bar: 2~cm. (c) Photograph of mouse with device implanted subcutaneously in the back. When the ambient light is off, the LED is visible through the skin. (d) Two paths of motion over 30~s of free movement tracked by video. (e) Power harvested by the device over the two paths. The efficiency of the energy transfer is $\eta= 0.75 \times 10^{-3}$ ($\pm0.14\times 10^{-3}$ standard deviation shaded).}
\end{figure*}

\begin{figure*}[b]
\centering
\includegraphics[width=17cm]{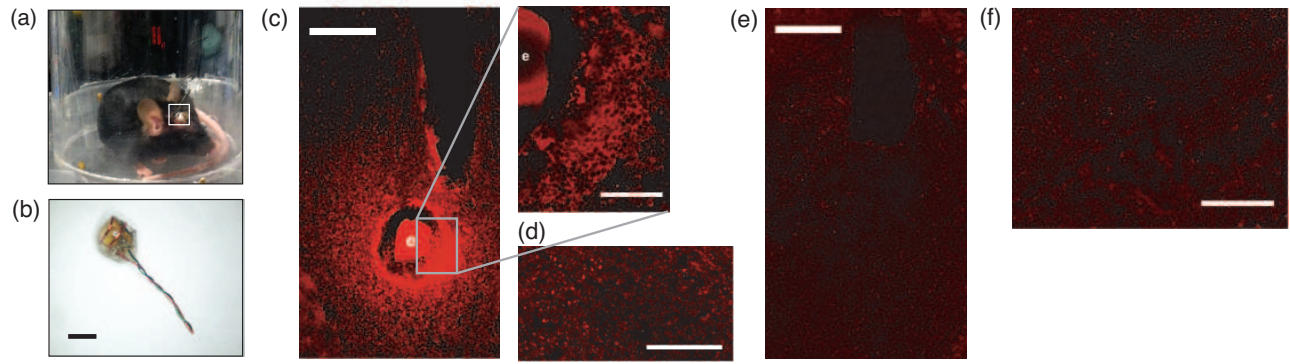}
\caption{Activation of wirelessly powered cranial electrodes induces robust c-Fos expression in the regions immediately around the electrode. (a) Photograph of mouse with implanted stimulator before the skin is sutured over the device. (b) Photograph of the device. Scale bar: 2~mm.(c) Intense c-Fos activation (red) in the region around electrode (labelled ``e'') inferior to electrode track. Scale bar: 200 $\mu$m. Inset region at 63$\times$ showing c-Fos activation in more detail. Scale bar: 100 $\mu$m. (d) c-Fos activation in region with electrode-related scarring, but without electrode track. Scale bar: 200 $\mu$m. (e) Sham electrode implantation and stimulation does not induce c-Fos expression near the electrode insertion site or in (f) regions with electrode-related scarring, but no electrode track. Scale bars: 200 $\mu$m.}
\end{figure*}

\end{document}